**Inner ear morphology in wild vs laboratory house mice**

Sabrina Renaud [1], Léa Amar [1], Pascale Chevret [1], Caroline Romestaing [2], Jean-Pierre Quéré [3], Corinne Régis [1], Renaud Lebrun [4]

[1] Laboratoire de Biométrie et Biologie Evolutive (LBBE), UMR 5558, CNRS, Université Claude Bernard Lyon 1, Université de Lyon, F-69622, Villeurbanne, France

[2] Laboratoire d'Ecologie des Hydrosystèmes Naturels et Anthropisés (LEHNA), UMR 5023, CNRS, ENTPE, Université Claude Bernard Lyon 1, Université de Lyon, F-69622, Villeurbanne, France

[3] Centre de Biologie et Gestion des Populations (INRA/IRD/Cirad/Montpellier SupAgro), Campus International de Baillarguet, Montferrier-sur-Lez Cedex, France

[4] Institut des Sciences de l'Évolution (ISE-M), UMR 5554, CNRS/UM/IRD/EPHE, Université de Montpellier, F-34090, Montpellier, France

**Short title**

Inner ear of wild vs lab mice

**ORCID ID**

Sabrina Renaud: 0000-0002-8730-3113

Léa Amar:

Pascale Chevret: 0000-0002-4186-875X

Caroline Romestaing: 0000-0002-6877-9626

Jean-Pierre Quéré:

Corinne Régis:

Renaud Lebrun: 0000-0002-5819-2653

**Abstract**

The semicircular canals of the inner ear are involved in balance and velocity control. Being crucial to ensure efficient mobility, their morphology exhibits an evolutionary conservatism attributed to stabilizing selection. Release of selection in slow-moving animals as been argued to lead to morphological divergence and increased inter-individual variation.

In its natural habitat, the house mouse *Mus musculus* moves in a tridimensional space where efficient balance is required. In contrast, laboratory mice in standard cages are severely restricted in their ability to move, which possibly reduces selection on the inner ear morphology. This effect was tested by comparing four groups of mice: several populations of wild mice trapped in commensal habitats in France; their second-generation laboratory offspring, to assess plastic effects related to breeding conditions; a standard laboratory strain (Swiss) that evolved for many generations in a regime of mobility reduction; and hybrids between wild offspring and Swiss mice. The morphology of the semicircular canals was quantified using a set of 3D landmarks and semi-landmarks analyzed using geometric morphometric protocols. Levels of inter-population, inter-individual (disparity) and intra-individual (asymmetry) variation were compared.

All wild mice shared a similar inner ear morphology, in contrast to the important divergence of the Swiss strain. The release of selection in the laboratory strain obviously allowed for an important and rapid drift in the otherwise conserved structure. Shared traits between the inner ear of the lab strain and domestic pigs suggested a common response to mobility reduction in captivity. The lab-bred offspring of wild mice also differed from their wild relatives, suggesting plastic response related to maternal locomotory behavior, since inner ear morphology matures before birth in mammals. The signature observed in lab-bred wild mice and the lab strain was however not congruent, suggesting that plasticity did not participate to the divergence of the laboratory strain.

However, contrary to the expectation, wild mice displayed slightly higher levels of inter-individual variation than laboratory mice, possibly due to the higher levels of genetic variance within and among wild populations compared to the lab strain. Differences in fluctuating asymmetry levels were detected, with the laboratory strain occasionally displaying higher asymmetry scores than its wild relatives. This suggests that there may indeed be a release of selection and/or a decrease in developmental stability in the laboratory strain.

**Introduction**

Locomotion is a crucial aspect of behavior, ensuring, among other things, access to trophic resources and escape from predators. Among the osteological structures related to locomotion, the inner ear (or bony labyrinth) (Fig. 1A) plays a key role by insuring balance. The inner ear is composed of the cochlea, which is involved in hearing, and of the vestibular system, including, in jawed vertebrates, three semicircular canals (SCCs) and a vestibule. The vestibular system is responsible for maintaining the balance of the body in relation to the movements performed during locomotion and as such, SCC morphology has been interpreted as related to locomotor behavior (Billet et al., 2012; Perier et al., 2016). Accordingly, in slow-moving animals such as sloths, selection related to locomotion is relaxed, leading to morphological divergence and increased inter-individual variance compared to fast-moving animals (Billet et al., 2012). Following the same line of reasoning, morphological differences in SCC morphology observed in domesticated animals such as pigs have been interpreted as due to mobility reduction (Evin et al., 2022). The processes and timing related to such responses remain hypothetical, especially with regard to the involvement of a plastic response to mobility reduction, that has been shown to have a pervasive impact on other osteological characters (Harbers et al., 2020; Neaux et al., 2021). In mammals, the inner ear develops early along ontogeny, with an almost adult morphology already achieved at birth (Costeur et al., 2017; Jeffery & Spoor, 2004), making it less prone to plastic variation than other bony features. Plastic variations of the inner ear morphology may however be driven by mobility patterns of the mother during the development of the embryos (Ronca et al., 2008).

The house mouse (*Mus musculus*) has a long history of association with humans, becoming commensal even before the Neolithic (Weissbrod et al., 2017). Being unintentional fellow-travelers in human exchanges allowed house mice to be one of the most successful invaders worldwide (Cucchi, 2008; Jones et al., 2012). In the resulting commensal habitat, "wild" mice have to forage in complex environments including three-dimensional exploration of buildings. By modifying exploratory behavior (Frynta et al., 2018), such habitat presumably exerts a strong selection on inner ear morphology, to ensure fast and agile locomotion. However, due to its ease of breeding in laboratory conditions, the house mouse has also been submitted to drastic changes in the regime of selection related to mobility: for about one century, laboratory strains (Wade et al., 2002; Yang et al., 2011) have been continuously bred in captivity. Compared to the ancestral 3D moving behavior, the usual laboratory cages presumably correspond to a regime of mobility reduction as much or greater than for livestock such as domestic pigs.

The aim of the present study is to take advantage of the house mouse ecology, allowing the comparison of wild-trapped and lab-bred animals, to test for short- and long-term effects of reduced mobility on inner ear morphology. Four groups of mice will be considered. (1) Wild-trapped commensal mice, which lived in natural conditions and were hence submitted to strong selection for efficient locomotion. (2) Laboratory offspring of wild mice. These mice have a wild genetic background but were raised in captivity for one or two generations, allowing the test of plastic response of the inner ear morphology, including maternal effects. (3) Typical laboratory mice of the Swiss strain. In this outbred strain, genetic diversity has been maintained, but mice were raised in captivity for decades. (4) Hybrids between wild-derived offspring (group 2) and Swiss (group 3). For all mice, the geometry of the inner ear morphology, with a focus on the semicircular canals, were quantified using a set of landmarks and sliding semi-landmarks (Evin et al., 2022). The following hypotheses were tested.

(1) Over generations, captivity should relax selection on the inner ear morphology and allow the Swiss strain to diverge from wild mice. In contrast, maintained selective pressures should limit divergence between wild populations. (2) By modifying growth conditions, captivity may trigger plastic differences between wild-trapped and laboratory-bred mice with a similar genetic background. (3) Neutral divergence due to drift and relaxation of adaptive pressures related to locomotion should lead to different patterns of hybridization. Neutral divergence, due to the accumulation of mutations in the two parental conditions, should lead to intermediate hybrids. In contrast, relaxation of selection may lead to the accumulation of unfavorable, recessive alleles in the lab strains, resulting in the dominance of the wild phenotype in hybrids. (4) Relaxed selection on inner ear morphology should contribute to increased inter-individual variance (disparity) (Billet et al., 2012; Perier et al., 2016) and increased intra-individual variance (asymmetry) (Lebrun et al., 2021) in lab-bred mice.

**Material**

*Sampling*

The material is composed of several sets of wild-trapped and laboratory-bred house mice (sampling is summarized in Table 1).

(1) Western European house mice (*Mus musculus domesticus*) were trapped in three commensal contexts. A population was collected during three trapping campaigns in a horse stable in Balan, nearby Lyon, France (45°49'09"N, 5°05'41"E) in 2015, 2017 and 2018. The mice considered in the morphometric study were sacrificed at capture. Two additional wild-trapped populations were documented. One of these populations corresponds to mice trapped in 2011-2012 in two neighboring farms in Tourch (Brittany, France): a large piggery in Kerloyou, and a traditional farm in the nearby place Kerc'hoaler, 1.8 km away. The other population was trapped in a roe deer enclosure in Gardouch, close to Toulouse (south-western France) in 2003-2004 (Renaud et al., 2017b).

(2) Laboratory offspring were obtained from a subset of mice trapped in Balan. In 2017, 31 animals were brought to the animal facility (ACSED, Lyon University). After about two months of acclimation, they were paired to obtain F1 offspring that were bred with standard rodent pellets (SAFE A04), with food and water *ad libitum*. Some of these F1 mice were paired to obtain F2 descendants (Savriama et al., 2022). Breeding was conducted in accordance with animal care guidelines. No ethical approval was required since no experiments were performed on the mice.

(3) Adult Swiss mice (Hsd:ICR [CD-1®]) were bought from Envigo. This strain was chosen because among classical laboratory strains, it is phylogenetically the closest to *Mus musculus domesticus* (Yang et al., 2011).

(4) To provide a tentative insight into hybridization effects, two mixed pairs were formed: a Swiss male x a F1 female, and a Swiss female x a F1 male. Two offspring of each pairs were included in the morphometric analysis. They were bred in the same conditions as the Swiss lab mice and offspring from wild-trapped mice.

Body weight was available for all mice except hybrids. Age was available for all Balan laboratory offspring and Swiss mice. Sex was available for most wild and lab mice.

All mice were killed in accordance with the European Parliament directive 2010/63/UE on the protection of animals used for scientific purposes. Tissues were collected and preserved in ethanol and/or minus 20°C for genetic analysis. Gardouch and Tourch specimens are stored at the Small Mammal Collection, Centre de Biologie pour la Gestion des Population, University of Montpellier, CIRAD, INRAE, Institut Agro, IRD, Montpellier, France (https://doi.org/10.15454/WWNUPO). Balan, Swiss and Hybrid individuals are stored as dissection specimens at the Laboratoire de Biométrie et Biologie Evolutive, University Lyon 1, Villeurbanne, France.

6| Pop | Code | SubGp | | N IE | N genet | N hap |
|---|---|---|---|---|---|---|
| Balan Wild | Bal_W | Bal_W1 | 2015 | 4 | 9 | 1 |
| | | | 2017 | 2 | 2 | 1 |
| | | Bal_W2 | 2018 | 11 | 13 | 2 |
| Balan Lab | Bal_L | Bal_L | | 21 | 10 | 1 |
| Gardouch | Ga | Ga | | 8 | $1^1$ | 1 |
| Tourch | To | ToKY | Kerloyou | 5 | $18^2$ | 2 |
| | | ToKH | Kerc'hoaler | 5 | $8^2$ | 1 |
| Swiss | SW | SW | | 14 | 2 | |
| Hybrids | Hyb | Hyb_fBL | Hybrid f BL x m SW | 2 | NA | |
| | | Hyb_fSW | Hybrid f SW x m BL | 2 | NA | |

Table 1. Sampling of the study, with definitions of the groups used in the different tests. N IE: number of specimens measured for the inner ear morphology. N genet: number of D-loop sequences, N hap: number of D-loop haplotypes. The Table includes data from this study, and from Bonhomme et al., 2011 ($^1$, Toulouse, near Gardouch), and Renaud et al., 2017b ($^2$).

**Methods**

*Extraction of 3D surfaces and (semi-)landmarks acquisition*

Skulls of wild mice from Balan, Tourch, Gardouch, most laboratory offspring and five Swiss were scanned at a cubic voxel resolution of 12 µm on the General Electric (GE) Nanotom microtomograph (µCT) of the AniRA-ImmOs platform of the SFR Biosciences, Ecole Normale Supérieure (Lyon, France). The dataset was complemented by one Balan Wild scanned at 12 µm, eight Balan Lab scanned at 17 µm, and nine Swiss scanned at 19 µm at the Mateis laboratory (INSA, Lyon, France), using a similar equipment.

The bony labyrinths were subsequently segmented using a two-step approach (Evin et al., 2022). A pre-segmentation of one slice every ~5-10 slices was performed using Avizo 2021.1 (Thermo Fisher Scientific). The Biomedisa smart interpolation tool (Lösel et al., 2020) was used to complement this pre-segmentation. The extracted left and right bony labyrinths were then exported as surface PLY files.

A series of landmarks and sliding semi-landmarks were digitized on the semicircular canals (SCC) of the left labyrinths or mirrored right labyrinths using MorphoDig (Lebrun, 2018).

The lateral SSC was described by 23 semi-landmarks anchored by two landmarks. The anterior and posterior SCCs were each described by 19 semi-landmarks anchored by two landmarks, and the common crus was described by 3 semi-landmarks anchored by two landmarks. One landmark, at the intersection of the common crus, the anterior and the posterior SCCs anchored the three corresponding curves, resulting in a total of 64 semi-landmarks and 6 landmarks per labyrinth (Fig. 1B).

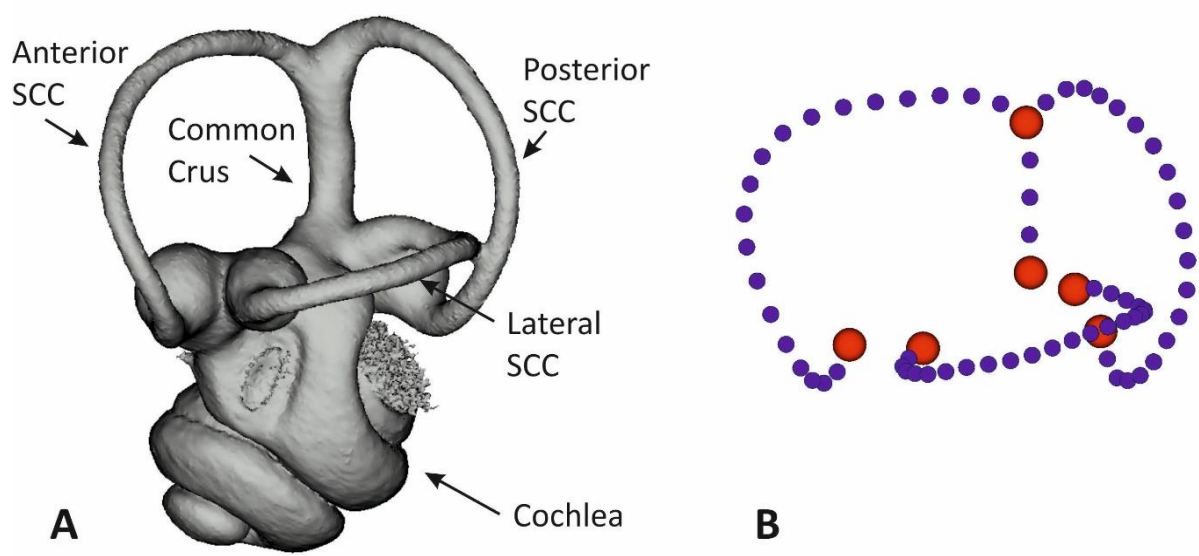

Figure 1. Example of the extracted surface of a left inner ear (Balan Wild #15). B. Corresponding set of landmarks (in red) and sliding semi-landmarks (in blue).

All pairs of inner ear surfaces are deposited in MorphoMuseum (Renaud et al., 2024). Information on the scanned specimens can be found in Supp. Table 1.

*Size estimates*

Body size. - Body weight was available for all mice except the hybrids. Since weight is more related to volume than to linear size, its cubic root (CRweight) was considered in the analyses.

Mandible size. – For each mouse, a 3D model of the mandible was also segmented, in most cases considering the right mandible. The 3D length of the mandible (MdL) was measured as the distance between the landmark located at posteriormost extremity of the condyle and the landmark at the anteriormost edge of the mandibular bone along the incisor lingual side. This measure was chosen because it can be assessed even on damaged mandibles, and because mandible size has been shown to display very little sexual dimorphism and to correlate well with body size (head + body length) (Renaud et al., 2017b). Landmarks were positioned using MorphoDig (Lebrun, 2018).

Inter-ear distance. – The distance between the two inner ears of a mouse (dRL) was calculated as the distance between the centroids of the right and left configurations of landmarks and semi-landmarks.

*Geometric morphometrics*

Coordinates were superimposed with a Generalized Procrustes Analysis (GPA) using the bending energy as the criterion to optimize semi-landmark superimposition. Two GPAs were performed: a first one on left / mirrored-right average configurations to study inter-individual variation, and one on both the left and mirrored-right configurations to assess intra-individual variation (asymmetry). The GPAs were performed using the R package geomorph (Adams & Otarola-Castillo, 2013). The aligned coordinates (Procrustes residuals) were thereafter considered as shape variables. The centroid size (Csize) of the left / mirrored-right average configuration provided an estimate of inner ear size for each mouse.

*Statistical analyses*

Differences between populations were tested using the following groups: Balan Wild (Bal_W, N = 17), Balan Lab (Bal_L, N = 21), Swiss (SW, N = 14), Hybrids (HYB, N = 4), Gardouch (Ga, N= 8) and Tourch (To, N = 10). Finer categories within these groups (years of trapping in Balan Wild; direction of cross for the hybrids; farms in Tourch) were graphically displayed to check for the consistency of the main groups.

Regarding the univariate size estimates, differences between groups were tested using non-parametric Kruskal–Wallis (KW) tests complemented by pairwise Wilcoxon tests. Linear regressions and Pearson correlations were used to test for covariation between univariate variables. Linear models were used to test for the effect of multiple factors and their interaction.

The shape variance was summarized using a Principal Component Analysis (PCA) applied to the variance-covariance matrix of the aligned coordinates. This variance **T** can be decomposed into two components: the between-group matrix **B** and the within-group matrix **W**, with **T** = **B** + **W**. A between-group PCA (later bgPCA) corresponds to the eigenanalysis of **B**; it was used to display differences between group means on synthetic axes.

Procrustes ANOVAs on the aligned coordinates were used to investigate the effect of various factors on inner ear geometry. Tests were performed on the total sampling using the population as factor, and separately within the best sampled populations (Balan Wild and Balan Lab). Pairwise differences between populations were tested on the Euclidean distances between groups, based on a permutational approach using the Morpho package (Schlager, 2017).

According to additive effects, hybrids should be in a strictly intermediate position between the parental strains. Any deviation from this prediction corresponds to a transgressive signal [e.g. (Nolte & Sheets, 2005)]. The degree of transgression of hybrids can thus be estimated as the sum of the distances between the hybrids and each one of the two parental strains compared to the distance between the two parental groups, expressed in percentage of the distance between parental groups (Renaud et al., 2012), i.e. [d(Hybrid, Parent1) + d(Hybrid, Parent2) - d(Parent1, Parent2)] / d(Parent1, Parent2) * 100.

Similarly, dominance of the parent 1 morphotype can be estimated by assessing the distance between hybrids and parent 1 with respect to the average distance between the hybrids and the two parents (Renaud et al., 2012): [d(Hybrid, Parent1) + d(Hybrid, Parent2)]/2 – d(Hybrid, Parent1), expressed as the percentage of the average distance of hybrids to the parents [d(Hybrid, Parent1) + d(F1 Hybrid, Parent2)]/2. Positive values then indicate closeness to parent 1, and negative values closeness to parent 2. Here, parent 1 corresponds to Balan laboratory offspring and parent 2 to the Swiss strain.

Morphological disparity was estimated as the Procrustes variance within each population, and differences between disparity levels were tested using permutational pairwise comparisons using the geomorph package (Adams & Otarola-Castillo, 2013).

Asymmetry between the right and left sides of a paired structure in a bilateral organism can be decomposed into consistent differences between the right and left sides (directional asymmetry, DA) and randomly fluctuating differences (fluctuating asymmetry, FA). While DA can be due to adaptive evolution [e.g. (Palmer, 2004)], FA is interpreted as due to developmental perturbations generating deviations from the perfect symmetry [e.g. (Alibert et al., 1994; Auffray et al., 1999)]. To assess this intra-individual variation, a Generalized Procrustes Analysis (GPA) was performed on the left and mirrored-right semi-landmark configurations following the same procedure as above. The shape variation was then decomposed into inter-individual variation, inter-side variation (directional asymmetry), and variation due to an individual x side interaction (fluctuating symmetry) (Adams & Otarola-Castillo, 2013). These components were statistically evaluated using Procrustes ANOVA. The Individual signed asymmetry index (signed.AI) provided a measure of shape asymmetry for each specimen (Klingenberg & McIntyre, 1998). Differences in FA levels between populations were tested using Kruskal-Wallis tests and pairwise Wilcoxon tests.

When relevant, probabilities were based on 9999 permutations. All analyses were performed under R (R Core Team, 2018). The geometric morphometric datasets are available as Supplementary Tables 2 (disparity: left / mirrored-right average configurations) and 3 (asymmetry: left and mirrored-right configurations).

*Genetic analyses*

DNA was extracted from ethanol-preserved or frozen tissue of Balan Wild, Balan Lab and Swiss mice (Table 1), using the DNeasy Blood and Tissue kit (Qiagen, France) or the Nucleospin 96 tissue kit (Macherey-Nagel, Germany) following the manufacturers' instructions.

The D-loop was amplified using previously described primers and protocol (Hardouin et al., 2010) and sequenced using the PCR primers by Biofidal (Vaux-en-Velin, France). The generated sequences were visualized and analysed using CLC Main Workbench (Qiagen, France). Sequences were aligned with MUSCLE implemented in SeaView 4 (Gouy et al., 2010). The 36 new sequences were deposited at EMBL under the accession numbers OY800699-OY800734.

This dataset was completed with 422 D-loop sequences from French localities obtained from previous studies (Belheouane et al., 2020; Bonhomme et al., 2011; Linnenbrink et al., 2013; Renaud et al., 2017b).

D-loop haplotypes were determined separately for the new and published data with DnaSP v6 (Rozas et al., 2017). The best substitution model (HKY+I+G) was determined with jModelTest (Darriba et al., 2012) using the Akaike criterion (Akaike, 1973). The phylogenetic tree was reconstructed using maximum likelihood with PhyMl 3.0 (Guindon et al., 2010) and Bayesian inference with MrBayes v3.2 (Ronquist et al., 2012) using this model. Node robustness was estimated with 1,000 bootstrap replicates with PhyMl and posterior probability with MrBayes. Markov chain Monte Carlo analyses were run independently for 2,000,000 generations with one tree sampled every 500 generations. The burn-in was determined graphically with Tracer v1.7 (Rambaut et al., 2018). We also checked that the effective sample sizes (ESSs) were above 200 and that the average SD of split frequencies remained <0.05 after the burn-in threshold. 10% of the trees were discarded and the resulting tree was visualized with FigTree v1.4 (Rambaut, 2016).

**Results**

*Phylogenetics*

Four haplotypes were identified among the 36 new sequences: Balan_2015-2017 (Balan Wild of 2015 and 2017), Balan 2018a (all but one mouse collected in 2018, and the 10 sequenced Balan Lab), Balan 2018b (one mouse collected in 2018) and Swiss (Table 1). As no DNA was available for the mice collected in Gardouch, sequences of the nearby Toulouse (Bonhomme et al., 2011) were used instead. The D-loop sequences of Tourch were retrieved from previous work (Renaud et al., 2017b) (Table 1). The new sequences and the data retrieved from previous work constituted a complete dataset of 107 haplotypes and 874 sites.

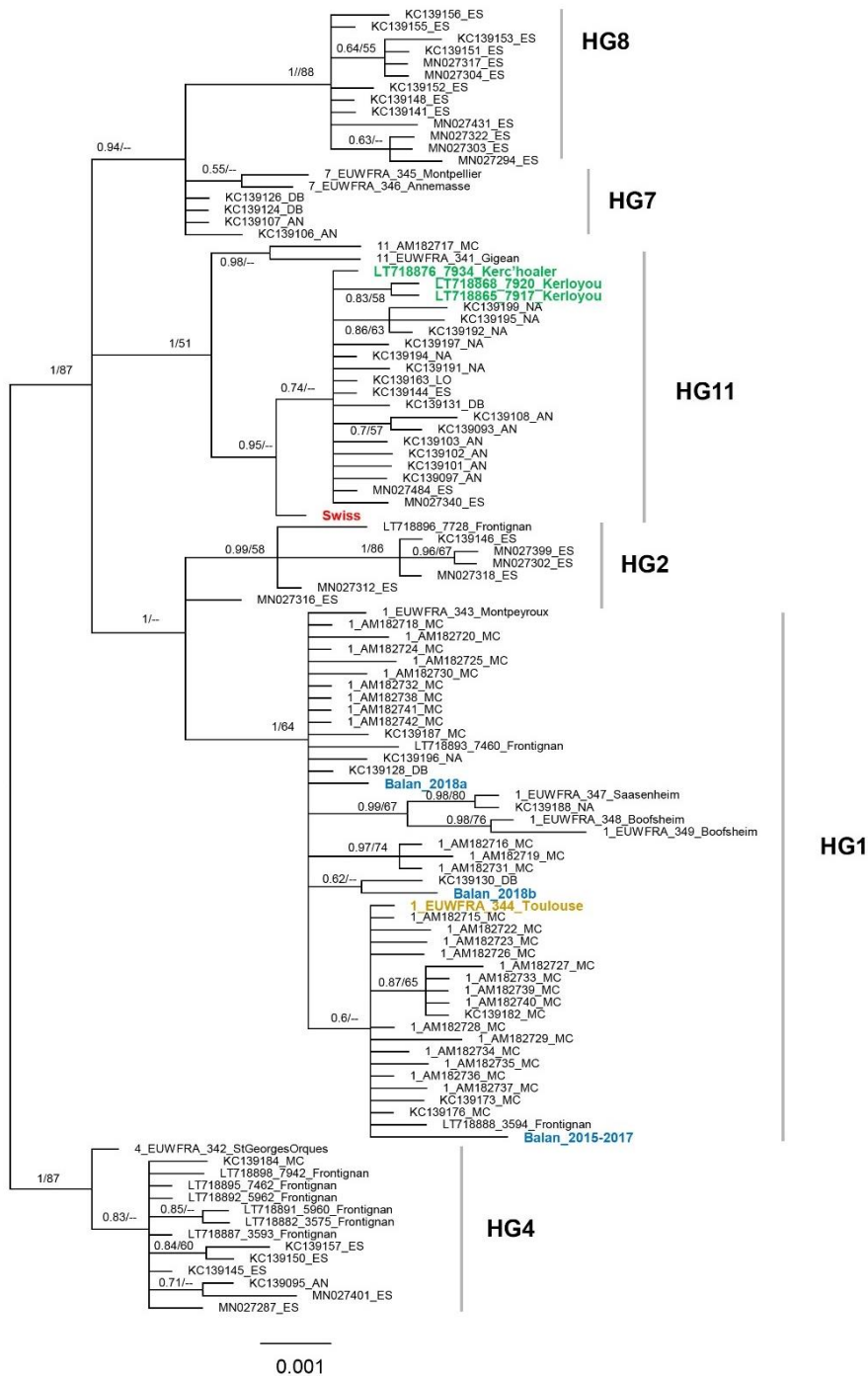

Figure 2. Bayesian phylogeny reconstructed with the D-loop haplotypes. The support of the nodes is indicated as follow: Posterior Probability (MrBayes)/Bootstrap Support (PhyMl). When the bootstrap support was below 50%, "/--" was indicated next to the posterior probability. The different haplogroups (sensu Bonhomme et al., 2011) are indicated next to the phylogeny. The haplotypes of the populations of interest are shown in bold and blue (Balan), light brown (Gardouch), green (Tourch) and red (Swiss).

The Bayesian phylogeny (Fig. 2) shows that the sequences Tourch and Balan belong to different haplogroups, respectively identified as HG11 and HG1 (Bonhomme et al., 2011). Within Tourch, mice from Kerloyou and Kerc'hoaler present a uniform haplotypic composition within each farm, slightly differing between the two neighboring spots. The Swiss strain presents a haplotypic signature belonging to the same haplogroup as Tourch mice (HG11).

Balan mice all belong to the same clade (HG1), but with a difference in composition over the years. Mice trapped and sacrificed in 2015 and 2017 have the same haplotype, whereas mice trapped in 2018 present another haplotype, with the exception of one mouse (Bal_W #24) displaying a third haplotype. Genotyped parents of the laboratory offspring present both dominant haplotypes, showing that they co-occurred in the population at the moment of trapping in 2017.

*Body, mandible and inner ear size*

Mandible size was related to the cubic root of body weight, without influence of the grouping (MdL ~ CRweight * population: CRweight $P < 0.0001$, population $P = 0.1993$, interaction $P = 0.0623$). A Pearson correlation of mandibular size vs cubic root of body weight further indicated a tight correlation between both variables ($R = 0.8979$). Therefore, mandible length was considered as a proxy for the size reached by the mouse at the time of sacrifice.

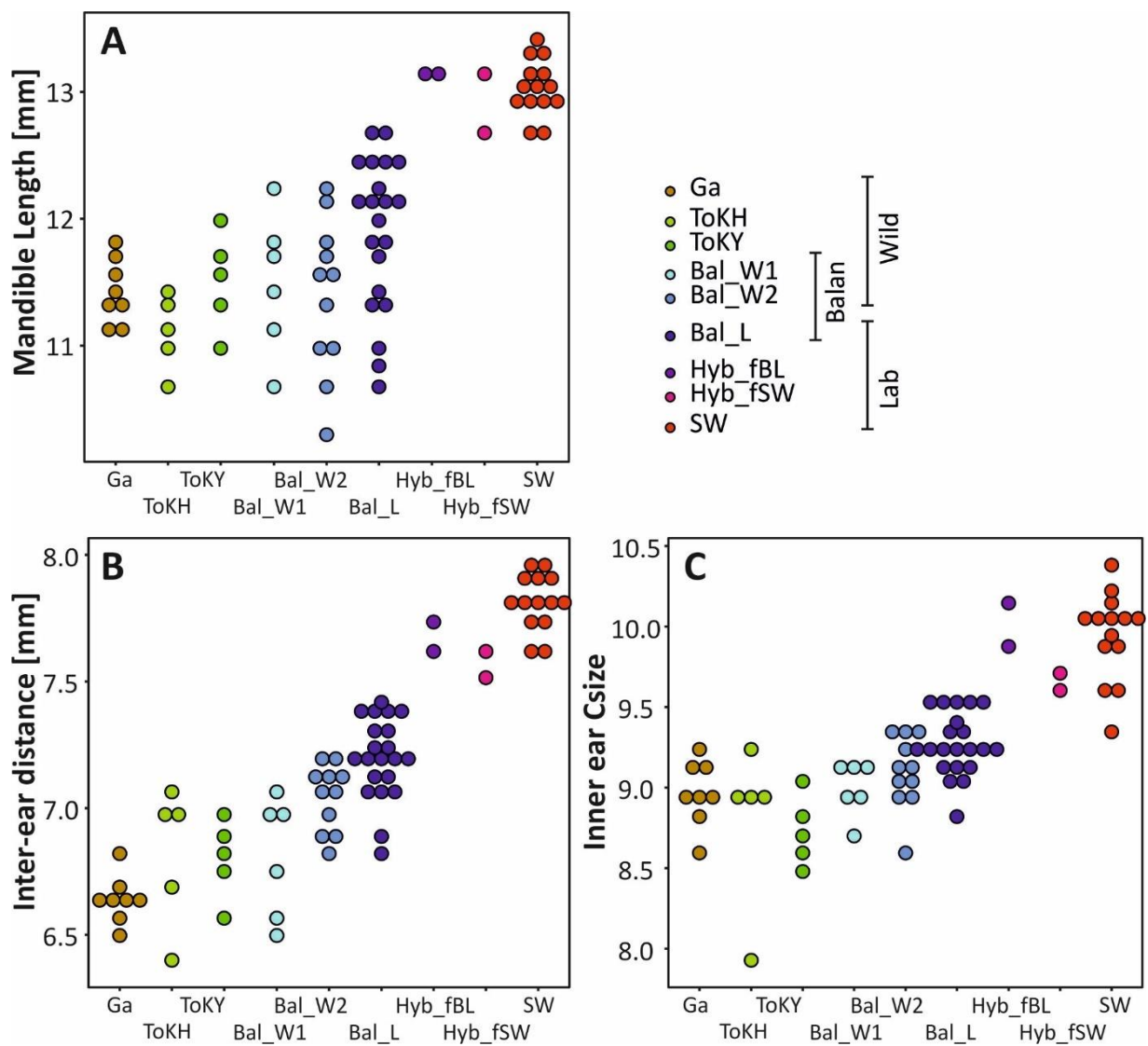

Figure 3. Size variation among populations. A. Mandible length. B. Inter-ear distance. C. Inner ear size, estimated by the centroid size (Csize) of the set of landmarks and sliding semi-landmarks).

The populations differed in mandible size (Fig. 3A, Table 2). This was mostly due to the larger size of the Swiss strain compared to wild mice. Hybrids were as large as their Swiss parents. Wild-trapped populations did not differ among each other, and did not differ from the Balan lab offspring.

Regarding the inter-ear distance (Fig. 3B, Table 2), the Swiss strain displayed larger values than all other populations, including hybrids which had nevertheless more widely spaced ears than wild-trapped mice and their lab-bred offspring. The wild-trapped populations (Balan, Tourch and Gardouch) did not differ among each other. The Balan laboratory offspring had more distant ears than wild-trapped animals.

A similar pattern was found when considering the inner ear centroid size (Fig. 3C, Table 2). Swiss mice had much larger ears than all other populations except hybrids. Balan laboratory offspring had slightly larger ears than wild-trapped populations, with similar ear size.

For all of these three size variables, the sub-populations (years of trapping in Balan, neighboring farms in Touch) displayed similar ranges of variations. The two directions of hybrid crosses did not differ much, and if at all, the hybrids with the wild mother displayed slightly larger values than the hybrids with the Swiss mother (Fig. 3).

| **MdL** | Swiss | Hybrid | Balan_Lab | Balan_Wild | Gardouch |
|---|---|---|---|---|---|
| Hybrid | 1 | - | - | - | - |
| Balan_Lab | **0.0000** | *0.0035* | - | - | - |
| Balan_Wild | **0.0000** | *0.0035* | 0.1245 | - | - |
| Gardouch | **0.0001** | *0.0323* | 0.1245 | 1 | - |
| Touch | **0.0000** | *0.0180* | 0.0618 | 1 | 1 |
| **dRL** | Swiss | Hybrid | Balan_Lab | Balan_Wild | Gardouch |
| Hybrid | *0.0105* | - | - | - | - |
| Balan_Lab | **0.0000** | *0.0014* | - | - | - |
| Balan_Wild | **0.0000** | *0.0023* | *0.0020* | - | - |
| Gardouch | **0.0001** | *0.0121* | **0.0000** | *0.0068* | - |
| Touch | **0.0000** | *0.0100* | **0.0001** | 0.0868 | 0.0868 |
| **Csize** | Swiss | Hybrid | Balan_Lab | Balan_Wild | Gardouch |
| Hybrid | 0.9455 | - | - | - | - |
| Balan_Lab | **0.0000** | *0.0016* | - | - | - |
| Balan_Wild | **0.0000** | *0.0030* | *0.0192* | - | - |
| Gardouch | **0.0001** | *0.0202* | *0.0098* | 0.9455 | - |
| Touch | **0.0000** | *0.0140* | **0.0003** | 0.0939 | 0.9455 |

Table 2. Tests of size differences between populations. MdL: mandible length. dRL: inter-ear distance. Csize: centroid size of the inner ear (R-L average configuration). Non-corrected probabilities of pairwise Wilcoxon texts are provided (H0: no difference between populations). In italics P < 0.05, in bold P < 0.001.

The inter-ear distance was significantly impacted by mandible size and population (dRL ~ MdL * population: MdL P < 0.0001, population P < 0001, interaction P = 0. 8951), but the populations shared parallel slopes, as indicated by the non-significant interaction term. The Swiss strain, having larger mandibles and more distant ears, mostly drove this correlation (Fig. 4A). Within each population, a significant relationship was found only in Bal_L (Pearson correlation: P = 0.0433, R = 0.4450).

A similar pattern was found between mandible size and inner ear centroid size (Csize ~ MdL * population: MdL P < 0.0001, population P < 0001, interaction P = 0.9161), but no significant relationship was found within any populations (Fig. 4B).

The inter-ear distance and inner ear centroid size were more tightly related (Fig. 4C). The group still had an effect on this relationship, but less so than in the relationship with mandible size (Csize ~ dRL * population: dRL P < 0.0001, population P = 0.0036, interaction P = 0.0106). The relationship was still valid within most well-sampled populations (Pearson correlations: Bal_W P = 0.0141; Bal_L P = 0.0020; Tourch P = 0.0017), but with differing slopes as indicated by the significant interaction term in the model "Csize ~ dRL * population".

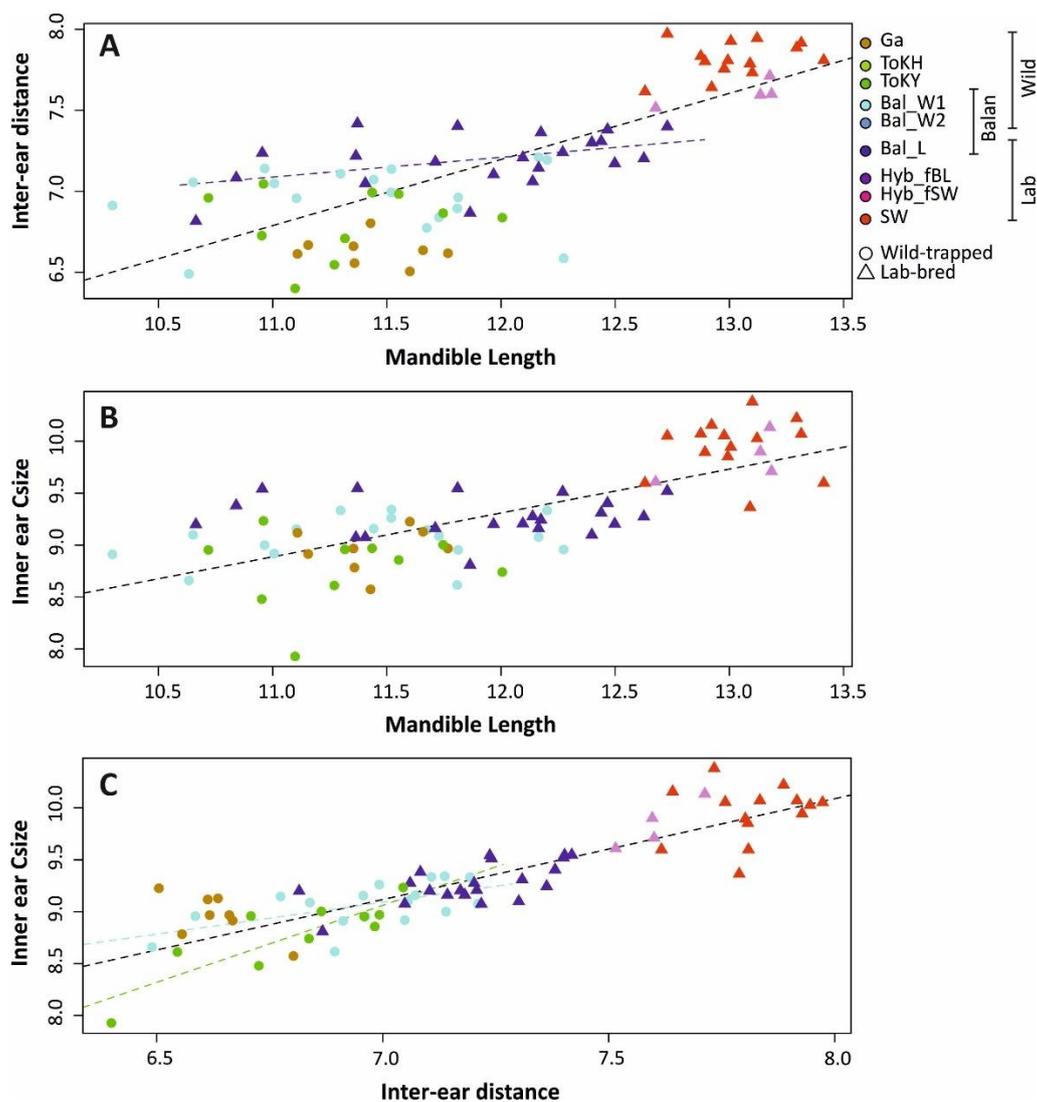

Figure 4. Relationship between size-related variables. A. Mandible size (MdL) vs inter-ear distance (dRL). B. Mandible size vs inner ear centroid size. C. Inter-ear distance vs inner ear centroid size. Lines correspond to significant Pearson correlations (non-significant correlations are not represented). In black, total sample. In colors, the corresponding population.

Models including sex as a co-factor with the population showed no significant effect of sex on the different relationships between size-related variables.

*Shape variation between populations*

Regarding inner ear shape, a model including population and sex indicated significant differences between populations but not between sexes (ProcD.lm: cords ~ pop * sex, Ppop < 0.0001; Psex = 0.1025; interaction P = 0.6382). Since neither size nor shape indicated the presence of sexual dimorphism, males and females were pooled in all subsequent analyses.

Patterns of inner ear shape variation were first investigated on a PCA (Fig.5A). The first axis (28.1% of the total variation) clearly isolated the Swiss strain towards positive scores, without any overlap with all wild mice (wild-trapped and lab-bred offspring). Towards the opposite side of the axis, the Balan laboratory offspring reached the most extreme scores. The Balan wild-trapped mice plot towards the center of the axis, with the different years occupying a similar range of variation. The Tourch group share the same shape space, with both farms plotting together. In contrast, the Gardouch population plots within the range of the Balan lab offspring, without displaying as extreme values along PC1.

According to this visual interpretation, the Swiss strain was highly differentiated from all wild-derived populations (Table 3). The Balan laboratory offspring differed from their wild relatives, but not from the Gardouch population. Some differences occurred among the wild-trapped populations, Gardouch and Tourch being different from Balan.

The shape differences are distributed all over the ear (Fig.5B). In Swiss mice compared to Balan mice trapped in the wild (Bal_W), the posterior canal appears expanded, the anterior canal contracted, and the lateral canal twisted, leading to an unbalanced shape with an expanded posterior part. In contrast, when Balan lab offspring are compared to their wild-trapped relatives, the anterior and posterior canals appear to be contracted dorso-ventrally, and the lateral canal appears straighter.

The few hybrids tended to be intermediate between both parental groups, with a low degree of transgression (19.2%) and a moderate dominance towards the Balan parents (22.4%). However, hybrids seemed to differ depending on the direction of the cross. Hybrids with the wild mother were half way to the Swiss strain (transgression = 31.6%, dominance towards Bal_L = 0.5%), whereas hybrids with the Swiss mother were closer to their wild relatives (transgression = 41.2%, dominance towards Bal_L = 31.5%).

|  | Balan_Lab | Balan_Wild | Gardouch | Hybrid | Swiss | Touch |
|---|---|---|---|---|---|---|
| Balan_Lab | – | 0.0211 | 0.0158 | 0.0262 | 0.0566 | 0.0280 |
| Balan_Wild | *0.0107* | – | 0.0252 | 0.0270 | 0.0457 | 0.0276 |
| Gardouch | 0.5651 | *0.0317* | – | 0.0325 | 0.0555 | 0.0246 |
| Hybrid | 0.1679 | 0.1507 | 0.0788 | – | 0.0413 | 0.0323 |
| Swiss | **0.0001** | **0.0001** | **0.0001** | *0.0028* | – | 0.0424 |
| Touch | *0.0034* | *0.0048* | 0.0946 | 0.0601 | **0.0001** | – |

Table 3. Shape differences between populations. Above the diagonal, Euclidean distances between group means. Below the diagonal, pairwise tests of shape differences between populations. Uncorrected permutational probabilities based on Euclidean distances between group means are provided. In italics P < 0.05, in bold P < 0.001.

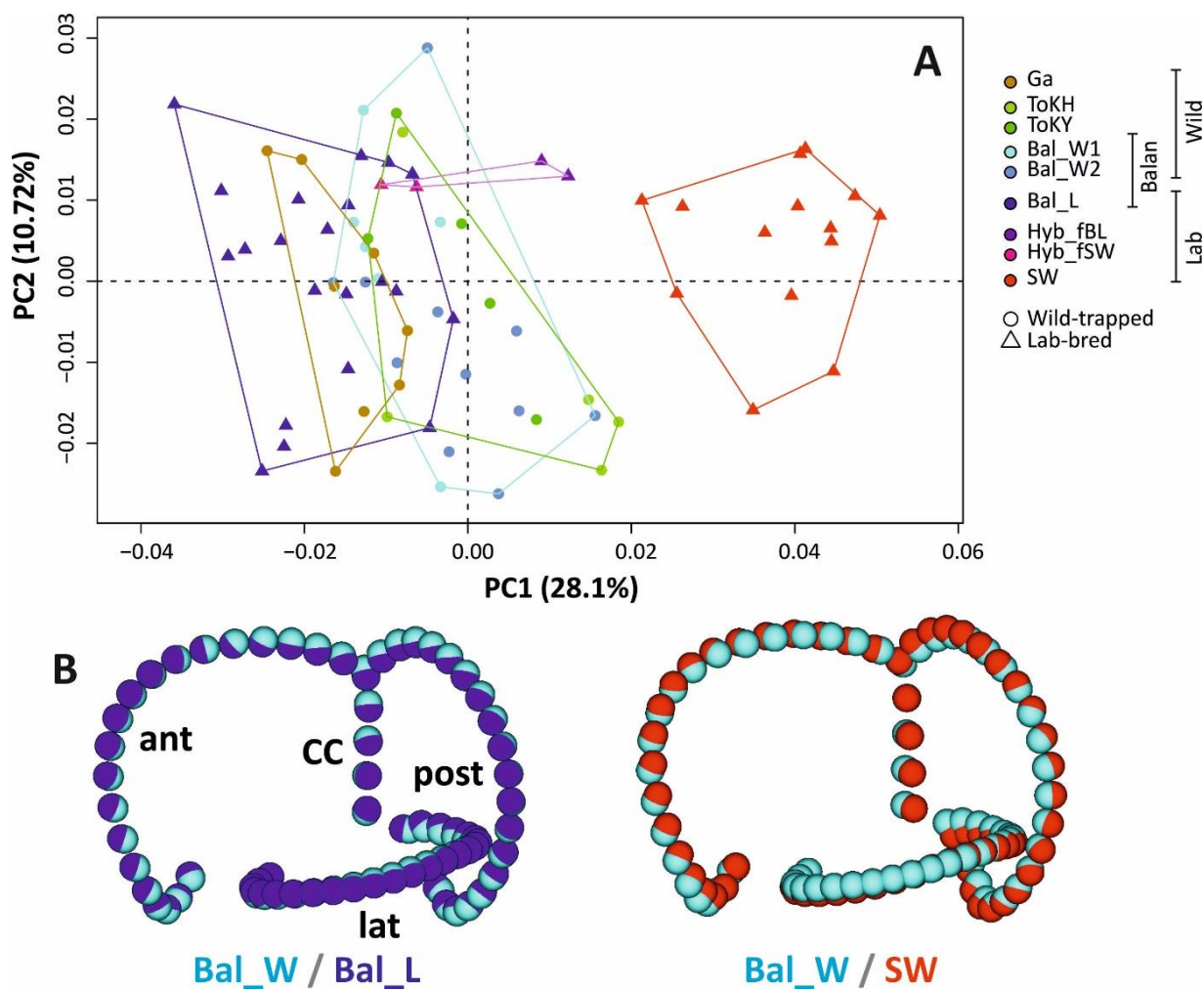

Figure 5. Patterns of shape variation of the inner ear. A. PCA on the aligned coordinates. Convex hulls enclose populations; the colors represent the sub-groups. B. Visualization of the mean shape of Bal_W (light blue) vs Bal_L (dark blue) and Bal_W vs SW (red).

A bgPCA indicated that including sub-structures within the main groups (Fig. 6) represented 38.1% of the total variance, vs. 31.9% for the main groups. When looking at the corresponding morphospace (Fig. 6), the different years from Balan appear very close. Some differences occurred between the two farms in Touch.

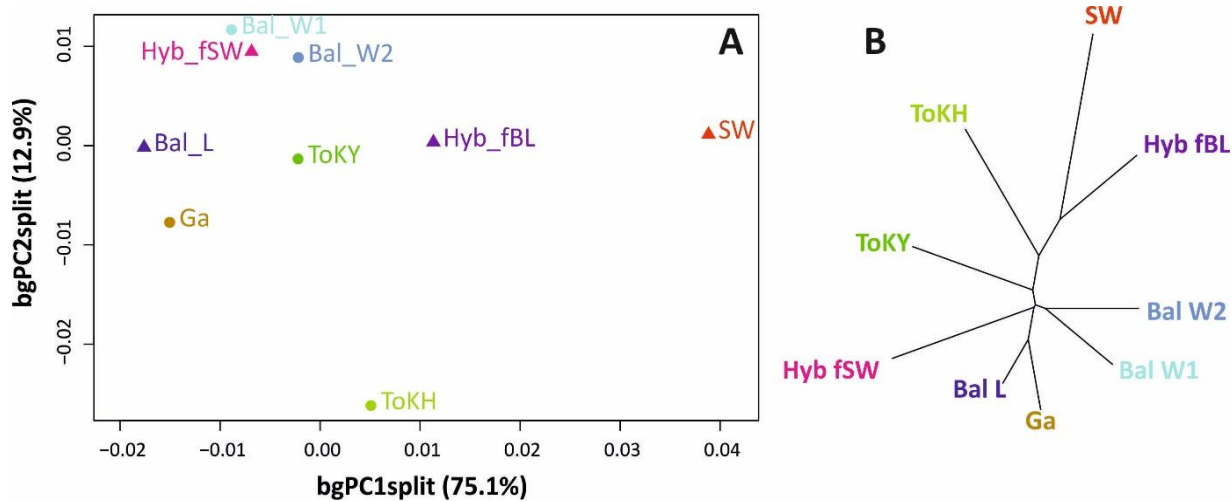

Figure 6. A. Mean shapes of the sub-groups on a between-group PCA. B. Unrooted neighbor-joining tree on the Euclidean distances between sub-group means.

*Relationship with size-related variables*

Inner ear shape was overall significantly related to mandible size, inter-ear distance and inner ear centroid size (ProcD.lm P < 0.0001; Fig.7). Yet, these relationships were mostly driven by the Swiss being larger in size and different in shape. Accordingly, population had a more important effect on inner ear shape than size-related variables (Table 4).

|  |  | Fac1 |  | Fac2 |  | Fac1:Fac2 |  |
|---|---|---|---|---|---|---|---|
| ProcD.lm |  | R² | P | R² | P | R² | P |
| Coords | MdL*pop | 8.9% | **< 0.0001** | 23.6% | **< 0.0001** | 5.5% | 0.2491 |
|  | dRL * pop | 12.4% | **< 0.0001** | 21.1 | **< 0.0001** | 6.6% | *0.0238* |
|  | CSize * pop | 11.8% | **< 0.0001** | 22.7 | **< 0.0001** | 6.5% | *0.0238* |
| Coords Bal_L | Age | 2.4% | 0.9759 |  |  |  |  |
|  | MdL | 3.4% | 0.8284 |  |  |  |  |
|  | dRL | 11.1% | *0.0027* |  |  |  |  |
|  | CSize | 7.4% | 0.0890 |  |  |  |  |
| Coords Bal_W | MdL | 5.1% | 0.6898 |  |  |  |  |
|  | dRL | 6.7% | 0.3663 |  |  |  |  |
|  | CSize | 7.6% | 0.2260 |  |  |  |  |

Table 4. Relationship between inner ear shape, size variables and population. R-squared and probabilities of a ProcD.lm model are provided. Coords: aligned coordinates; MdL: mandible length; CSize: inner ear centroid size; dRL: inter-ear distance. In italics P < 0.05, in bold P < 0.001.

Furthermore, when considering regression scores derived from the regression of size vs aligned coordinates as variables describing the allometric signal, in most cases the relationship was not valid within populations (Fig. 7). In the case of inner ear centroid size only, the two wild populations of Gardouch and Touch displayed relationships congruent with the general between-group trend (Fig. 7C).

To further explore this issue, the relationship was explored in the two Balan groups, for which a large range of age / growth stage is covered. In the Balan laboratory offspring, neither age, mandible size nor inner ear centroid size influenced inner ear shape (Table 4). Only the inter-ear distance had a significant effect ($R^2$ = 11.1%). When considering Balan wild-trapped mice, none of the size-related variables had a significant effect on shape (Table 4).

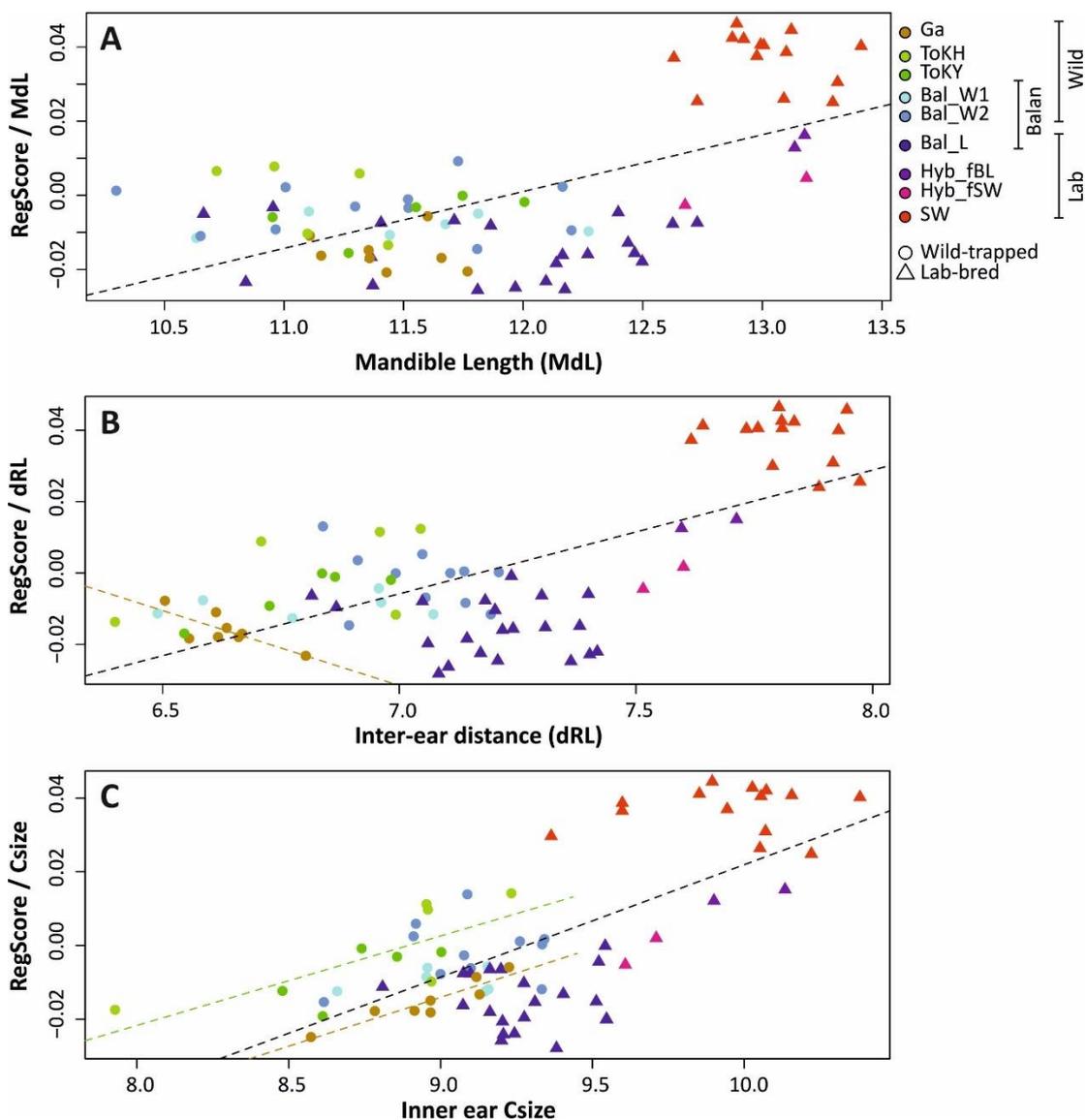

Figure 7. Relationships between inner ear shape and size-related variables. Inner ear shape is described by regression scores of aligned coordinates vs each size-related variable. A. Shape vs mandible size. B. Shape vs inter-ear distance. C. Shape vs inner ear centroid size. In black, the regression on the total sample. In color, significant regressions within populations.

*Disparity among populations*

Morphological disparity was evaluated in the different populations (Fig. 8; Table 5). The wild-trapped populations of Balan and Tourch were significantly more variable in shape than lab-bred groups and Gardouch.

| Variance |  |  |  |  |  |  |
|---|---|---|---|---|---|---|
|  | Swiss | Hybrid | Balan_Lab | Balan_Wild | Gardouch | Tourch |
| Swiss | 0 |  |  |  |  |  |
| Hybrid | 0.000339 | 0 |  |  |  |  |
| Balan_Lab | 0.000059 | 0.000398 | 0 |  |  |  |
| Balan_Wild | 0.000376 | 0.000715 | 0.000317 | 0 |  |  |
| Gardouch | 0.000109 | 0.000448 | 0.000050 | 0.000267 | 0 |  |
| Tourch | 0.000489 | 0.000829 | 0.000431 | 0.000114 | 0.000380 | 0 |
| P-Values |  |  |  |  |  |  |
|  | Swiss | Hybrid | Balan_Lab | Balan_Wild | Gardouch | Tourch |
| Swiss | 1 |  |  |  |  |  |
| Hybrid | 0.1719 | 1 |  |  |  |  |
| Balan_Lab | 0.6964 | 0.0974 | 1 |  |  |  |
| Balan_Wild | *0.0165* | *0.0063* | *0.0268* | 1 |  |  |
| Gardouch | 0.5778 | 0.0971 | 0.7860 | 0.1589 | 1 |  |
| Tourch | *0.0078* | *0.0027* | *0.0091* | 0.5227 | 0.0691 | 1 |

Table 5. Disparity levels and differences in the different populations. Above, pairwise absolute differences between variances. Below, probabilities of significant differences. In italics P < 0.05, in bold P < 0.001.

*Asymmetry levels among populations*

A GPA was performed on both sides of each specimen (left inner ear and mirror image of the right inner ear). A Procrustes ANOVA was performed on the resulting aligned coordinates, with side and individual as factors to assess the relative importance of systematic differences between side (directional asymmetry, DA) and of fluctuating asymmetry (FA). For inner ear centroid size, FA was significant ($P_{ind}$ = 0.0001) whereas DA was not ($P_{side}$ = 0.2729). Regarding inner ear shape, FA was highly significant ($P_{ind}$ = 0.0001) with some component of DA ($P_{side}$ = 0.0011).

A Procrustes ANOVA on the FA component of shape variation showed significant differences between populations (ProcD P = 0.0077). A Kruskall-Wallis test on the FA index (signed.AI) also showed differences between populations (P = 0.0349). The results suggested that the Swiss and hybrid groups are overall not more asymmetric than wild populations, but they occasionally display the most asymmetric specimens (Fig. 8B).

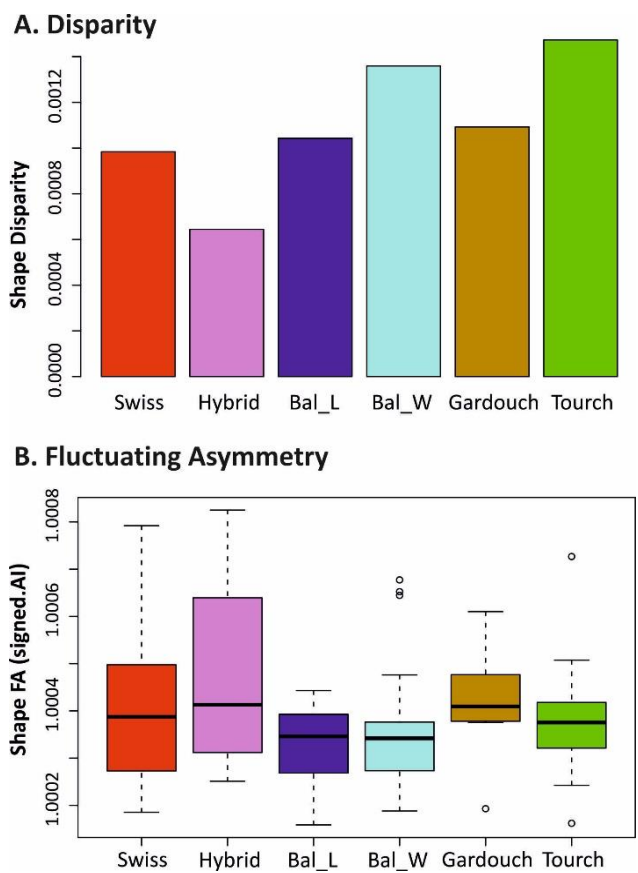

Figure 8. Disparity and shape asymmetry of inner ear shape as a function of the populations. A. Disparity level in the different population. B. Fluctuating asymmetry (signed. AI).

**Discussion**

*Differences in inner ear morphology between wild- and captive-bred mice*

Three commensal populations of house mice were compared. Although all were trapped in France, the most distant were separated by more than 700 km. For Tourch and Balan, genetic data are available ((Renaud et al., 2017b), this study), showing that mice belong to different haplogroups. Despite these genetic differences, the three commensal populations display limited differences in inner ear morphology, including both size and shape. Local genetic differences between farms (Tourch Kerloyou vs Kerc'hoaler) were associated with small differences in ear shape. Despite a genetic turnover in Balan from 2015 to 2018, inner ear size and shape remained stable. These results show that inner ear morphology can vary to a small degree between wild populations (Evin et al., 2022). Stabilizing selection on wild populations may be acting to limit divergence; conversely, environmental differences might locally modulate developmental conditions and selective pressures.

24Gardouch mice, fostering in a roe deer enclosure, were probably less exposed to 3D vertical movements than Balan and Tourch mice, living in buildings.

In contrast, the Swiss strain was characterized by an important morphological differentiation, with no overlap with the wild-trapped mice. Such an important difference is usually characteristic of inter-specific or even inter-generic divergence (e.g. (Perier et al., 2016)). Two ranges of non-mutually exclusive explanations can be involved. First, the Swiss strain was founded one century ago from a small number of mice (Aldinger et al., 2009). Genetic variation was subsequently maintained by large effective population size, but random fixation of alleles nevertheless occurred, resulting in a genetic signature comparable to insular populations (Rice & O'Brien, 1980). Such an evolutionary context is known to promote fast and pronounced morphological evolution (e.g. (Chevret et al., 2021)). Consistently, compared to wild relatives, the Swiss strain is characterized by important differences in other characters such as body size, and molar morphology (Savriama et al., 2022). Second, the Swiss strain evolved over generations in a regime of strong mobility reduction that relaxed stabilizing selection on inner ear morphology, allowing for an important divergence in shape over a century.

Adaptive or functional explanations for the differences in semicircular canal shape remain often speculative. However, the expanded posterior part of the Swiss inner ear leads to an unbalanced morphology distorting the arrangement between the three semi-circular canals. Similarly, slow-moving primates display anterior and posterior canals of unbalanced size, although in favor of the anterior canal (Perier et al., 2016). The arrangement of the three semi-circular canals is assumed to be related to angular head velocities (fast head rotation) during locomotion (Malinzak et al., 2012; Perier et al., 2016). Altogether, this suggests that the Swiss inner-ear morphology might be maladaptive. Since the posterior and lateral canals reach their final morphology later than the anterior one (Costeur et al., 2017), this unbalanced shape might be at least partly the consequence of the large body size of Swiss mice.

A comparison with the inner ear morphological change in domestic pigs compared to wild boars shows some common features, with a contraction of the anterior canal and an expansion of the posterior one (Evin et al., 2022), suggesting a shared trend towards a "captive" inner ear morphology. Domestic pigs also show an increased inner-ear size compared to their wild relatives (Evin et al., 2022), thus supporting a possible role of allometric growth in this "captive" signature. Not mutually exclusive, it may be a component of the complex skull shape changes involved in the evolution of domesticated forms (Neaux et al., 2021), and possibly a component of the highly integrated "domestication syndrome" (Wilkins et al., 2014).

The short-term effect of captive breeding was less important than the divergence of the Swiss strain, but nevertheless significant. Lab-bred offspring of wild-trapped mice displayed inner ears of larger size and different shape compared to their wild-trapped relatives. This difference occurred over only one or two generations, and thus corresponds to a plastic response to breeding conditions, providing evidence that even a presumed conserved structure can vary in a plastic way. In the inner-ear morphospace, the lab-bred offspring of wild mice diverged in the opposite direction from Swiss mice, showing that no common "captive" signal was involved. This further suggests that the divergence of Swiss is not due to plastic variations, and that the evolution of their peculiar morphology was not facilitated by the trajectory of plastic response, according to a genetic assimilation process (Aubret & Shine, 2009; Badyaev, 2005).

This discrepant response to captive breeding may be due to the fact that it involves a relaxation of stabilizing selection, rather than a common constraint favoring convergent evolution. In laboratory-bred offspring of wild mice, the limitation of maternal movements might have caused a decreased stimulation on the developing inner ears (Ronca et al., 2008), leading to their dorso-ventrally compact shape.

*Sources of within-population variation in inner ear morphology*

The inner ear is a structure that develops early along ontogeny, reaching its adult size and shape before birth in mammals (Costeur et al., 2017; Jeffery & Spoor, 2004). This suggests that little or no variation should occur along post-natal growth of mice, despite a massive remodeling of the skull (e.g. (Zelditch et al., 2004)). In agreement, little influence of inner ear size and inter-ear distances, and no effect of body size, was found on inner ear morphology within populations. Prenatal growth may nevertheless modulate inner ear development and generate some intra-population variance, but proxies for adult body size poorly reflect size at birth and do not allow testing for such effects within populations.

In apparent contradiction, inner ear morphology was found to scale with body size at a large phylogenetic scale (Costeur et al., 2019), but such studies involve much larger size differences than observed within populations. Swiss mice may represent a case of intraspecific variation in body size important enough to impact inner ear morphology, since Swiss offspring are probably already larger than wild ones at birth. Yet, allometry alone cannot explain the divergence of the Swiss inner ear, since hybrids between wild and Swiss mice had body and inner ear sizes as large as their Swiss parents, but morphologies closer to the smaller, wild parent.

Sexual dimorphism has been suggested as another source of intraspecific variation of inner ear shape (Evin et al., 2022). In the present study, it was found to be of little or no influence on inner ear size and shape. This is probably related to the limited sexual dimorphism observed in house mice, including for osteological characters such as the mandible (Renaud et al., 2017b).

*A complex response to hybridization*

Despite the small set of hybrids between Swiss and wild-derived mice, they can provide a tentative insight into the impact of hybridization on inner ear morphology. The most striking feature was a discrepant pattern regarding size and shape of the inner ear. Inner ear size clearly displayed a dominance towards the large Swiss parent, as did mandible size and inter-ear distance. Dominance towards the size of the largest parent was repeatedly found in crosses between mouse strains, for different traits such as mandible (Renaud et al., 2012) or molar size (Renaud et al., 2017a; Savriama et al., 2022), suggesting a general pattern regarding the response of size to hybridization in such crosses.

As for inner ear shape, hybrids tended to be intermediate between both parents, with a moderate degree of transgression (~20%). Transgression can occur due to the accumulation of genes with antagonistic effects in each parental group, which can produce transgression from the F2 generation onwards (Rieseberg et al., 1999), and/or a modular response to hybridization leading to hybrid phenotypes differing from the mid-shape between both parents (Renaud et al., 2012). Such effects appear to be moderate here.

However, inner ear morphology was not homogeneous within hybrids, contrary to what has been observed for molar shape of the same mice (Savriama et al., 2022). Hybrids with the wild mother were midway from both parental morphologies, in agreement with an additive model, whereas hybrids with the Swiss mother displayed a pronounced dominance towards the wild phenotype. These results suggest maternal effect and/or different genetic/epigenetic effects depending on the direction of the cross. Additional sampling would be required to validate these results.

*Disparity and asymmetry in response to breeding conditions*

One of the initial hypotheses of the study was that relaxed selection in lab-bred mice should lead to increased levels of disparity and asymmetry (Billet et al., 2012; Lebrun et al., 2021; Perier et al., 2016). These predictions were not fulfilled, neither in the lab-bred offspring of wild mice nor in the Swiss strain. This shows that over one century of evolution in captivity, canalization has been maintained for inner ear morphology.

In fact, contrary to the expectation, mice bred in laboratory conditions showed decreased disparity levels compared to wild-trapped populations. This may be partly due to a reduced genetic diversity in the lab-bred groups. The Swiss strain is outbred, with pedigrees specifically designed to maximize genetic diversity and heterozygosity, nevertheless it has evolved as a closed system over many generations and accordingly, its disparity regarding molar shape appears to be reduced compared to wild mice (Savriama et al., 2022). The laboratory offspring of the wild mice descend from parents trapped while a genetic turnover occurred in the population, leading to mixed ancestry in the lab offspring, mitigating the issue of potential inbreeding. In contrast, wild house mouse populations are known to function in small demes, with groups of related individuals structured at a very small geographic scale (Pocock et al., 2004) but genetic variation is nevertheless documented in Balan and Tourch. Decreased genetic diversity in the lab-bred mice compared to wild populations may therefore contribute to their reduced disparity. Growth conditions in the laboratory are more homogeneous than in the wild, possibly further contributing to a reduced inter-individual variance.

Regarding intra-individual variance, the results suggested that inner ear shape was impacted by little directional asymmetry but some fluctuating asymmetry, although the results must be treated with caution as repeated measurements were not performed. Differences in the levels of fluctuating asymmetry were small, but some specimens from the Swiss strain and their hybrids displayed higher FA levels than occurring in wild populations, thus partly validating the initial hypothesis of an increased intra-individual variance in the lab-bred mice (Lebrun et al., 2021). However, laboratory offspring of the wild mice did not display increased levels of asymmetry, showing that this effect was not due to the breeding conditions *per se*, but rather to a long-term consequence of inbreeding and/or the absence of selection against asymmetric specimens in laboratory strains (Leamy, 1984; Soulé, 1967).

Mice from the laboratory strain therefore display a pronounced morphological differentiation from their wild relatives, and possibly decreased developmental stability. Such mice are used for applied research in balance and hearing (Ohlemiller et al., 2016), and yet may not adequately represent a wild phenotype. This echoes the growing awareness that laboratory strains are not being realistic models due to their long history of inbreeding and selection for laboratory conditions (Rosshart et al., 2017).

**Conclusions**

Inner ear morphology is assumed to be under strong stabilizing selection, due to its role in controlling balance and hence mobility. Accordingly, its evolution mostly occurs due to neutral evolution (Mennecart et al., 2022). The present study however provide evidence that a change in selective regime can lead to an important and rapid morphological divergence, then departing from a slow, neutral pace of differentiation (Renaud et al., 2007). The morphological signature observed in the house mouse laboratory strain shares traits with inner ears of domestic pigs (Evin et al., 2022) and even slow-moving primates (Perier et al., 2016), suggesting a possible common response to mobility reduction.

The expected increase in inter-individual disparity associated with relaxed selection was in contrast not validated in the present dataset, possibly due to confounding factors. The level of genetic diversity in wild populations may influence shape disparity, raising caution when interpreting differences between wild populations. Furthermore, the present study provided evidence of plastic variations in inner ear shape in relation with conditions of growth. This response is probably related to maternal movements since the inner ear reaches its adult morphology before birth. This opens new questions regarding the importance of locomotory behavior on the shape of the inner ear, when considering fine-scale intraspecific variations. House mice can foster in very different spatial structures, ranging from buildings with a high three-dimensional complexity to ground-dwelling populations in the case of feral mice. Such ecological differences can trigger rapid response in traits associated with nutrition, such as mandible shape (Souquet et al., 2019) or incisor dynamics (Renaud et al., 2019) due to the involvement of plastic responses. The present study suggests that even inner ear morphology might vary on such ecological time-scale. Further studies addressing patterns of inter-population differentiation would be required to go deeper into this fine tuning between behavioral adjustment to local ecological conditions and inner ear morphology.

**Acknowledgements**

Angéline Clair, Laetitia Averty and Julie Ulmann are warmly thanked for their investment in breeding the laboratory offspring in the animal husbandry unit ACSED. The horse stable Les Peupliers (Balan) is thanked for authorization and support during trapping. We also acknowledge the contribution of SFR Biosciences (UMS3444/CNRS, US8/Inserm, ENS de Lyon, UCBL) AniRa-ImmOs facility and the Mateis laboratory (INSA, Lyon, France), and we particularly thank Mathilde Bouchet and Justine Papillon for their kind assistance during the scanning sessions. Finally, we thank Samuel Ginot as well as the editor, Dr. Philip Cox, for their constructive remarks on the manuscript. This work was supported by the Fédération de Recherche BioEEnViS – FR3728 of University Claude Bernard Lyon 1.

**Statements and Declarations:** The authors declare they have no conflict of interest.

**Supplementary Material**

| SpecID | Population | Group | MdL [mm] | Weight [g] | Age [days] | Sex |
|---|---|---|---|---|---|---|
| Balan_02 | Balan_Wild | Commensal | 10.6 | 7.6 | | |
| Balan_04 | Balan_Wild | Commensal | 12.3 | 21.7 | | F |
| Balan_06 | Balan_Wild | Commensal | 11.8 | 17.9 | | M |
| Balan_07 | Balan_Wild | Commensal | 10.9 | 11.3 | | F |
| Balan_08 | Balan_Wild | Commensal | 11.7 | 17.3 | | F |
| Balan_11 | Balan_Wild | Commensal | 11.4 | 10.9 | | M |
| Balan_12 | Balan_Wild | Commensal | 11.1 | 11.7 | | F |
| Balan_BAL15 | Balan_Wild | Commensal | 11.0 | 11.2 | | M |
| Balan_BAL16 | Balan_Wild | Commensal | 11.5 | 15.9 | | M |
| Balan_BAL17 | Balan_Wild | Commensal | 10.7 | 7.9 | | M |
| Balan_BAL18 | Balan_Wild | Commensal | 10.3 | 9.4 | | F |
| Balan_BAL19 | Balan_Wild | Commensal | 11.7 | 16.3 | | M |
| Balan_BAL20 | Balan_Wild | Commensal | 11.3 | 12.7 | | M |
| Balan_BAL21 | Balan_Wild | Commensal | 12.2 | 17.6 | | M |
| Balan_BAL22 | Balan_Wild | Commensal | 11.5 | 13.9 | | F |
| Balan_BAL23 | Balan_Wild | Commensal | 11.8 | 15.4 | | F |
| Balan_BAL24 | Balan_Wild | Commensal | 12.2 | 14.4 | | M |
| Balan_BAL25 | Balan_Wild | Commensal | 11.0 | 9.8 | | M |
| Gardouch_3419 | Gardouch | Commensal | 11.2 | 13.3 | | M |
| Gardouch_3432 | Gardouch | Commensal | 11.4 | 13.4 | | F |
| Gardouch_3437 | Gardouch | Commensal | 11.4 | 15.3 | | M |
| Gardouch_3439 | Gardouch | Commensal | 11.1 | 12.4 | | M |
| Gardouch_3450 | Gardouch | Commensal | 11.4 | 17.4 | | M |
| Gardouch_3453 | Gardouch | Commensal | 11.7 | 14.2 | | F |
| Gardouch_3459 | Gardouch | Commensal | 11.8 | 12.9 | | M |
| Gardouch_3462 | Gardouch | Commensal | 11.6 | 12.3 | | M |
| Tourch_7819 | Tourch | Commensal | 12.0 | 12.0 | | F |
| Tourch_7821 | Tourch | Commensal | 11.7 | 11.0 | | F |
| Tourch_7839 | Tourch | Commensal | 11.0 | 9.0 | | M |
| Tourch_7873 | Tourch | Commensal | 11.6 | 13.0 | | M |
| Tourch_7877 | Tourch | Commensal | 11.3 | 11.0 | | F |
| Tourch_7922 | Tourch | Commensal | 11.1 | 15.0 | | M |
| Tourch_7923 | Tourch | Commensal | 11.3 | 16.0 | | F |
| Tourch_7925 | Tourch | Commensal | 11.4 | 13.0 | | M |
| Tourch_7927 | Tourch | Commensal | 11.0 | 9.0 | | F |
| Tourch_7932 | Tourch | Commensal | 10.7 | 9.0 | | M |
| Balan_Lab_167 | Balan_Lab | Lab_Offspring | 11.4 | 21.9 | 48 | M |
| Balan_Lab_17032205_01 | Balan_Lab | Lab_Offspring | 12.5 | 16.1 | 66 | F |
| Balan_Lab_188 | Balan_Lab | Lab_Offspring | 12.2 | 11.6 | 32 | M |
| Balan_Lab_192 | Balan_Lab | Lab_Offspring | 11.4 | 13.4 | 28 | M |
| Balan_Lab_194 | Balan_Lab | Lab_Offspring | 11.4 | 13.9 | 46 | M |
| Balan_Lab_196 | Balan_Lab | Lab_Offspring | 11.8 | 14.0 | 44 | F |
| Balan_Lab_30x17 | Balan_Lab | Lab_Offspring | 11.7 | 10.1 | 27 | M |
| Balan_Lab_319 | Balan_Lab | Lab_Offspring | 12.1 | 18.5 | 68 | F |
| Balan_Lab_325 | Balan_Lab | Lab_Offspring | 12.4 | 22.5 | 74 | M |

| SpecID | Population | Group | MdL | Body weight | Age | Sex |
|---|---|---|---|---|---|---|
| Balan_Lab_329 | Balan_Lab | Lab_Offspring | 12.4 | 21.0 | 74 | M |
| Balan_Lab_330 | Balan_Lab | Lab_Offspring | 12.5 | 23.6 | 74 | M |
| Balan_Lab_35 | Balan_Lab | Lab_Offspring | 11.9 | 14.7 | 98 | F |
| Balan_Lab_40x56 | Balan_Lab | Lab_Offspring | 10.7 | 9.2 | 24 | M |
| Balan_Lab_46 | Balan_Lab | Lab_Offspring | 12.1 | 16.0 | 85 | F |
| Balan_Lab_47x61 | Balan_Lab | Lab_Offspring | 11.0 | 10.8 | 22 | F |
| Balan_Lab_54 | Balan_Lab | Lab_Offspring | 12.3 | 17.4 | 73 | F |
| Balan_Lab_56 | Balan_Lab | Lab_Offspring | 12.0 | 16.3 | 74 | F |
| Balan_Lab_82 | Balan_Lab | Lab_Offspring | 12.2 | 25.8 | 118 | M |
| Balan_Lab_86 | Balan_Lab | Lab_Offspring | 12.7 | 24.0 | 108 | M |
| Balan_Lab_92 | Balan_Lab | Lab_Offspring | 12.6 | 21.0 | 112 | M |
| Balan_Lab_BB3weeks | Balan_Lab | Lab_Offspring | 10.8 | 9.0 | 21 | |
| hyb_125xSW_01 | Hybrid | Hybrid | 13.2 | | | F |
| hyb_125xSW_02 | Hybrid | Hybrid | 13.1 | | | F |
| hyb_SWx126_01 | Hybrid | Hybrid | 12.7 | | | F |
| hyb_SWx126_02 | Hybrid | Hybrid | 13.2 | | | F |
| SW0_348 | Swiss | Strain | 12.6 | 28.1 | 71 | |
| SW0bis_350 | Swiss | Strain | 12.9 | 27.7 | 73 | |
| SW0ter | Swiss | Strain | 13.0 | 34.4 | 64 | M |
| SW1 | Swiss | Strain | 13.4 | 44.3 | 87 | M |
| SW2 | Swiss | Strain | 13.1 | 39.9 | 118 | M |
| SW339 | Swiss | Strain | 13.0 | 39.4 | 72 | |
| SW341 | Swiss | Strain | 12.7 | 39.6 | 70 | |
| SW342 | Swiss | Strain | 12.9 | 40.3 | 70 | |
| SW343 | Swiss | Strain | 12.9 | 37.5 | 74 | M |
| SW345 | Swiss | Strain | 13.3 | 32.8 | 71 | |
| SW347 | Swiss | Strain | 13.1 | 29.5 | 73 | |
| SW5 | Swiss | Strain | 13.0 | 41.6 | 99 | M |
| SWF3 | Swiss | Strain | 13.1 | 41.5 | 103 | F |
| SWF4 | Swiss | Strain | 13.3 | 38.5 | 109 | F |

Supplementary Table 1. Sampling of the study. Identifier (SpecID), population and group for the different specimens, together with mandible length (MdL), body weight, age and sex.